\documentclass[10pt,twocolumn,superscriptaddress,nofootinbib,floatfix,aps,prb,citeautoscript,longbibliography]{revtex4-2}
\usepackage{float}

\usepackage[T1]{fontenc}
\usepackage{lineno}

\usepackage{graphicx}
\usepackage{color}
\usepackage{array}
\usepackage{dcolumn}
\usepackage{bm}
\usepackage{multirow}
\usepackage{chemformula}

\usepackage{booktabs}
\usepackage{tabularx}
\usepackage{comment}
\usepackage{tikz}
\usepackage{xstring}
\usepackage{lipsum}
\usepackage{subcaption}

\usepackage[abs]{overpic}
\usepackage{graphicx}
\usepackage{amsmath}
\usepackage{orcidlink}
\usepackage[capitalise]{cleveref}
\usepackage{enumerate}

\begin{document}
\newcommand{\bochum}{Research Center Future Energy Materials and Systems of the University Alliance Ruhr and Interdisciplinary Centre for Advanced Materials Simulation, Ruhr University Bochum, Universitätsstraße 150, D-44801 Bochum, Germany}
\newcommand{\DTU}{Computational Atomic-scale Materials Design, Department of Physics, Technical University of Denmark, DK-2800 Kongens Lyngby, Denmark}

\newcommand{\antoine}[1]{\textcolor{orange}{#1}}
\newcommand{\revision}[1]{\textcolor{blue}{#1}}
\newcommand{\toReplace}[1]{\textcolor{purple}{#1}}
\
\newcommand{\change}[1]{{#1}}

\makeatletter
\newcommand{\corrauthor}[1]{%
  \gdef\@corrauthor{{\par\centering\small\normalfont#1\par}}%
}
\pretocmd{\frontmatter@RRAPformat}{\@corrauthor}{}{}
\makeatother

\title{Accelerating point defect photo-emission calculations with machine learning interatomic potentials}

\author{Kartikeya Sharma$^*$\,\orcidlink{0009-0009-3300-6568}}
\affiliation{\DTU}
\author{Antoine Loew\,\orcidlink{0009-0008-5018-4895}}
\affiliation{\bochum}
\author{Haiyuan Wang\,\orcidlink{0000-0002-2411-1501}}
\affiliation{\DTU}
\author{Fredrik A. Nilsson\,\orcidlink{0000-0002-0163-3024}}
\affiliation{\DTU}
\author{Manjari Jain\,\orcidlink{0000-0002-2077-7381}}
\affiliation{\DTU}
\author{Miguel A. L. Marques\,\orcidlink{0000-0003-0170-8222}}
\affiliation{\bochum} 
\author{Kristian S. Thygesen\,\orcidlink{0000-0001-5197-214X}}
\affiliation{\DTU}
\email{thygesen@fysik.dtu.dk}
\corrauthor{$^*$Corresponding author: \texttt{kartsh@dtu.dk}}

\begin{abstract}
\begin{center}
\textbf{ABSTRACT}
\end{center}
We introduce a computational framework leveraging universal machine learning interatomic potentials (MLIPs) to dramatically accelerate the calculation of photoluminescence (PL) spectra of atomic or molecular emitters with \emph{ab initio} accuracy. By replacing the costly density functional theory (DFT) computation of phonon modes with much faster MLIP phonon mode calculations, our approach achieves speed improvements exceeding an order of magnitude with minimal precision loss. We benchmark the approach using a dataset comprising \emph{ab initio} emission spectra of 791 color centers spanning various types of crystal point defects in different charge and magnetic states. The method is also applied to a molecular emitter adsorbed on a hexagonal boron nitride surface. Across all the systems, we find excellent agreement for both the Huang-Rhys factor and the PL lineshapes. This application of universal MLIPs bridges the gap between computational efficiency and spectroscopic fidelity, opening pathways to high-throughput screening of defect-engineered materials. Our work not only demonstrates accelerated calculation of PL spectra with DFT accuracy, but also makes such calculations tractable for more complex materials.
\end{abstract}

\maketitle

\section{Introduction}
The study of defects in crystalline materials constitutes a cornerstone of condensed matter physics and materials science. The crystal defects can strongly influence, and sometimes even determine, the basic physical characteristics of the host material, including its magnetic, electronic, optical, and mechanical properties. Consequently, controlling the types and concentrations of defects is extremely critical for the development of advanced materials for a range of technological applications, including electronic devices, optical components, solar cells, and batteries~\cite{Wolfowicz2021, WANG201723, Sarkar2018, Jhuria2024}. 

Within the area of quantum technology, point defects in wide band gap semiconductors can act as a basis for single photon sources (e.g. the NV$^-$ defect in diamond), quantum sensors (e.g. via the optically detected magnetic resonance (ODMR) effect), or quantum information storage (e.g in the form of electron spins localised at the defect)~\cite{Jelezko2006, Marcus_2013}. In all these applications, the interaction with light is crucial for addressing and controlling the quantum state of the defect. 

One challenging aspect of the light-defect interaction is the strong coupling to the dynamic degrees of freedom of the host lattice, particularly under non-equilibrium conditions such as photo-excitation. Computational methodologies based on density functional theory (DFT) and many-body perturbation theory have emerged as indispensable tools for bridging atomic-scale defect configurations with macroscopic observables, enabling predictive design of defect-engineered materials~\cite{JONES1998287, DEAK2008187, Gali2023}. A central quantity in this regard is the photoluminescence (PL) spectrum of the defect, which encodes the interplay between electronic transitions on the defect and the quantized lattice vibrations, i.e., the phonons. The \emph{ab initio} calculation of the PL spectrum involves a ground state (GS) calculation to determine the equilibrium geometry of the defect, followed by an excited state (ES) calculation to obtain the relaxation pathway after photo-excitation. From the GS and ES configurations, one constructs a high-dimensional configuration coordinate vector, capturing the multi-dimensional displacement of atomic positions during electronic excitation (or de-excitation). This displacement vector is then expanded in terms of the phonon normal modes to yield the electron-phonon spectral function from which the PL spectrum can be derived~\cite{Alkauskas2014}. The calculation of the phonon modes of the many-atom supercell represents a bottleneck for \emph{ab initio} PL calculations.

Over the past decade, several methods leveraging artificial intelligence and big data have entered the field of materials science. Among those, machine learning interatomic potentials (MLIPs)~\cite{behler_perspective_2016, schmidt_recent_2019, graser_machine_2018, Unke2021} are perhaps the most significant. These techniques provide atomic energies and forces with \emph{ab initio} precision but orders of magnitude faster than DFT.  Particularly powerful are the recently introduced universal MLIPs based on graph neural networks. These models have been trained on sufficiently large data sets to be directly applicable across a broad variety of chemical compositions and structures without the need for retraining or fine-tuning~\cite{Chen2019, chen_universal_2022, batatia_mace_2023, Deng2023, Neumann2024, Park2024, Liao_2023, Choudhary2021}.  While universal MLIPs have shown remarkable capabilities for several standard tasks such as structure optimization, molecular dynamics, and the calculation of phonons in pristine crystals~\cite{Chmiela2023, Stocker2022, benchphonon}, their application to optical properties of defects remains unexplored. 

In this work, we present a machine learning (ML)-accelerated framework for calculating PL spectra of color centers by leveraging MLIPs for phonon predictions. After introducing the computational method, we benchmark seven universal MLIPs against a dataset of \emph{ab initio} Huang-Rhys (HR) factors for 791 point defects in different 2D host crystals and further validate their performance for a few bulk defect systems, namely the NV$^-$ center in diamond and two substitutional defects in silicon. From this benchmark, we identify Mattersim-v1 as the best-performing model for the current application. Interestingly, we find that the excellent performance of the method is independent of the charge and magnetic state of the defects, even though the MLIPs have no explicit knowledge of charges or magnetic moments. We further show that the excellent performance of the ML-framework on the HR factor prediction extends to the detailed lineshape of the PL spectra for both crystal point defects as well as solid-molecule interfaces.

\section{Results}
\label{sec:results}

\subsection{Predicting Huang-Rhys factors}
We first focus on the prediction of the important Huang–Rhys (HR) factor, $S$. The HR factor is a dimensionless parameter quantifying the strength of the electron-phonon coupling of an emitter and is widely used to characterize photoemission spectra. To determine $S$, we first evaluate the structural displacement vector, $\Delta\mathbf{R}$, connecting the electronic ground state and excited state geometries,
\begin{equation}
    \Delta R_{\alpha,i} = R_{\alpha,i}^{\mathrm{(ES)}} - R_{\alpha,i}^{\mathrm{(GS)}}
\end{equation}
where $\alpha$ denotes an atom and $i\in\{x,y,z\}$ its Cartesian coordinates. The atomic configurations of the ground state and excited states are obtained by relaxing the atoms in a supercell containing the defect using DFT with the electron occupation numbers corresponding to the ground state and an excited state of the defect, respectively (see Methods for more details). We then calculate the mass-weighted displacement vector, $ \Delta \mathbf Q$,
\begin{equation}
    \label{deltaQ}
    \Delta Q_{\alpha,i} = \sqrt{M_\alpha} \Delta R_{\alpha,i}
\end{equation}
where $M_\alpha$ is the mass of atom $\alpha$.

Next, to quantify the projection of the structural displacement onto individual phonon modes, we project $\Delta\mathbf{Q}$ onto the normalized phonon eigenvectors, $\mathbf{e}^{(k)}$. 
The mass-weighted displacement amplitude along phonon mode $k$ is given by 
\begin{equation} 
    \label{Qk}
    Q_k = \sum_{\alpha,i}\sqrt{M_\alpha} \, e_{\alpha i}^{(k)} \, \Delta R_{\alpha, i}
\end{equation}
It clearly holds that $\Delta Q^2 = \sum_{k}Q_k^2$.

The partial HR factor of mode $k$ is given by
\begin{equation}
    \label{partial_HR}
    S_k = \frac{\omega_k Q_k^2}{2\hbar}
\end{equation}
    where $w_k$ is the phonon frequency for the mode $k$. $S_k$ measures how much phonon mode $k$ is activated during the optical transition. The electron-phonon spectral function is defined as
\begin{equation}
    \label{SpectralFn}
    S(\omega) = \sum_k{S_k \delta(\hbar\omega-\hbar\omega_k)}.
\end{equation}
The total HR factor, $S$, is obtained by integrating $S(\omega)$ over all frequencies. Finally, the PL spectrum is obtained by applying a generating function to the electron-phonon spectral function~\cite{Alkauskas2014}.

The main bottleneck in the computational workflow described above is the calculation of the phonon modes and frequencies. This is because typical supercells used for defect calculations contain at least 100 atoms in order to ensure that the interaction between the periodically repeated images of the defect is negligible. Consequently, the determination of the phonons requires several hundred DFT calculations. Moreover, these DFT calculations are expensive as the defect supercell typically has no or only a few symmetries. Accelerating the calculation of photoemission from point defects should therefore focus on this phonon bottleneck. The strategy explored in the present work is to perform the entire phonon calculation using a universal MLIP.   
\mbox{}\\

\subsection{Machine learning models}
The seven MLIP models considered in the present work are listed in Table \ref{table:summaryHRerr}. The models were selected based on their performance on standard materials modelling tasks benchmarked by the Matbench leaderboard, as well as a recent benchmark study on phonons in perfect crystals~\cite{benchphonon, Riebesell_matbench}. Among the considered models, the M3GNet~\cite{Chen2019} and MACE-MP-0~\cite{batatia_mace_2023} were some of the earliest universal MLIPs. The MACE-MPA-0 is a further development of MACE-MP-0, utilizing a larger training set and with more parameters. The SevenNet-0~\cite{Park2024} is a scalable improvement of the NequIP~\cite{batzner_e3-equivariant_2022} model, which utilizes an equivariant graph neural network architecture. ORB-V2~\cite{Neumann2024} and eqV2-M~\cite{Barroso2024} are examples of non-conservative models, i.e., models that predict forces and energies separately rather than computing the forces as derivatives of the energy. Finally, Mattersim-V1-5M (MtS)~\cite{mattersim} was chosen as it is the best universal MLIP for the computation of phonons in pristine crystals~\cite{benchphonon}.

\begin{figure*}[ht!]
  \centering
  \includegraphics[width=\linewidth]{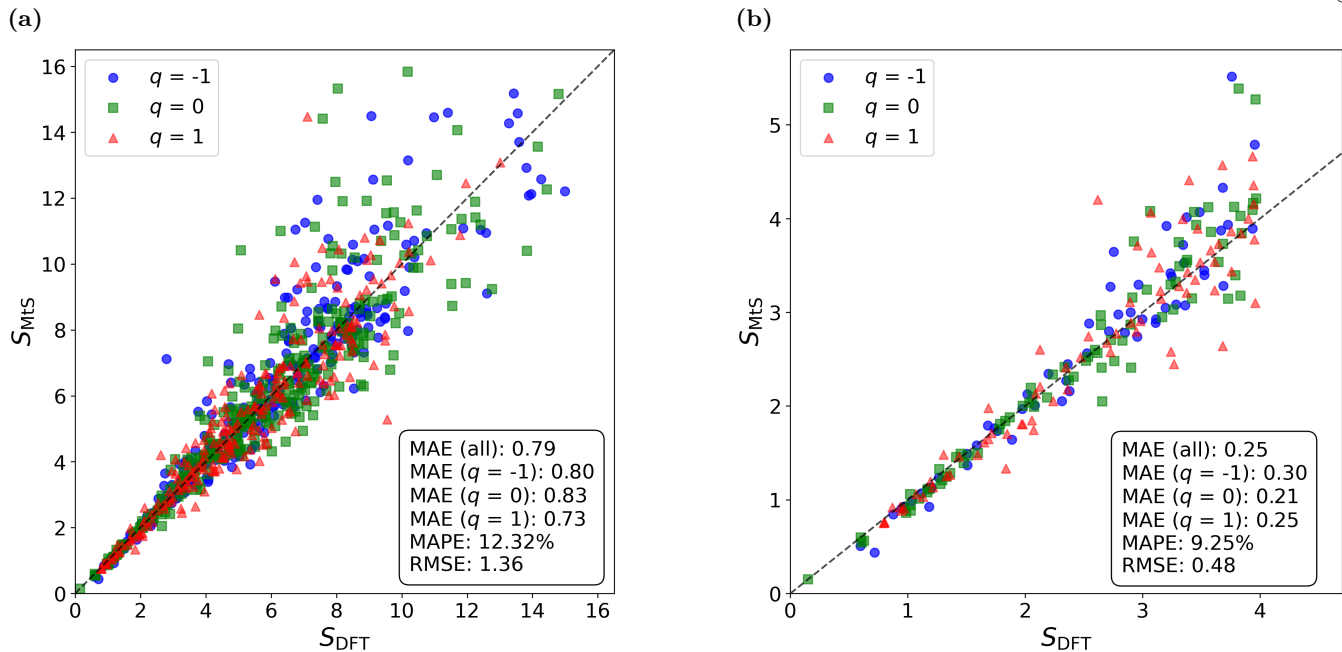}
  \caption{\textbf{Comparison of the HR factor calculated from MtS-predicted phonons versus the DFT values.} (a) shows the comparison across the full dataset, while (b) focuses on defects with $S_{\mathrm{DFT}} < 4$. The color of the markers denotes the charge state of the defect ($q = -1$, $0$, and $1$). The variance of the errors is 0.22 for lower HR factors ($S_{\mathrm{DFT}} < 4$), and rises to 2.41 for larger HR factors ($S_{\mathrm{DFT}} \geq 4$).}
  \label{fig:parity_s}
\end{figure*}

\subsection{Benchmark dataset}

For the benchmark, we used a dataset comprising a total of 791 point defects in the ten 2D host crystals: BN, GeSe, HfS$_2$, MoS$_2$, MoSe$_2$, MoTe$_2$, SnS$_2$, WS$_2$, WSe$_2$, and ZrS$_2$. 
The defects include vacancies as well as native and extrinsic substitutions. The extrinsic dopants comprise the atoms B, C, N, F, Al, Si, P, S, Cl, Sc, Ti, V, Cr, Mn, Fe, Co, Ni, Cu, Zn, Ga, Ge, As, Se, and Br. Double defects are created by combining two single-site defects on neighboring atomic sites (not all combinations of extrinsic dopants are included). All the defect supercells are large enough to ensure a minimum distance between periodically repeated defects of at least 15 \AA, and they contain around 100 atoms. For every defect, the charge states $+1$ and $-1$ are considered, if the corresponding charge transition levels lies within the band gap. The reported total of 791 defects includes such charge states.

The defect structures were relaxed with DFT-PBE, and the optical properties, including the lowest excitation energy, the HR factor, and the PL lineshape, were calculated. The DFT-based defect calculations were performed as part of a high-throughput project aimed at discovering novel spin defects for quantum technological applications\footnote{To be published}. 
For this reason, the dataset contains only defects with finite magnetic moments (non-magnetic defects were filtered out after the DFT relaxation step). In all the excited state calculations, the total spin is conserved, i.e., the spin is never flipped during an excitation. 

The current work employs the dataset of 2D defects, supplemented by a few bulk defects and a molecular emitter, to benchmark the proposed ML-based method. Note that the molecular benchmark system is only considered in its neutral charge state, which is non-magnetic, i.e., the ground state is a singlet. Therefore, to obtain accurate ZPL energies, we applied spin purification to correct for the open-shell nature of the excited-state. Following the procedure in Ref. \citenum{wang2025two}, the true singlet-state energy is approximated as $E = 2E_{\rm m} - E_{\rm t}$, where $E_{\rm m}$ and $E_{\rm t}$ are the energies of the spin-mixed state and the corresponding triplet states, respectively.

\subsection{Benchmark results}
\cref{table:summaryHRerr} summarizes the predictions of the seven MLIP models for the HR factors of our benchmark dataset. The reported values represent the difference between the predicted HR factor and the DFT results.

Two models, MACE-MPA-0 and MatterSim-v1-5M (MtS), are seen to perform significantly better than the others. Both models exhibit the lowest mean absolute error (MAE) and a reasonably low maximum error. However, MtS has a slightly lower MAE and RMSE, indicating better error variance compared to MACE-MPA-0. Additionally, SevenNet-0 and eqV2-M have the lowest maximum error, but have higher MAE. Based on these empirical observations, we focus on the MtS model in the rest of our study.

\begin{table}[htp]
\centering
\caption{Summary of the performance of different universal MLIPs for prediction of Huang-Rhys (HR) factors. The columns show the mean absolute error (MAE), the root mean squared error (RMSE), and the maximal error on a dataset comprising DFT calculated HR factors for 791 point defects in two-dimensional semiconductors and insulators.}
\label{table:summaryHRerr}
\begin{tabular*}{\columnwidth}{@{\extracolsep{\fill}} l c c c c}
\toprule
Model               & MAE ($\downarrow$)   & RMSE   & Max error \\
\midrule
Mattersim-V1-5M     & 0.8   & 1.4    & 12.7       \\
MACE-MPA-0          & 1.0   & 1.5    & 12.2      \\
M3GNet              & 1.2   & 1.9    & 17.8      \\
SevenNet-0          & 1.3   & 1.8    & 10.4      \\
MACE-MP-0           & 1.5   & 3.9    & 57.7      \\
eqV2-M              & 1.9   & 2.5    & 10.8      \\
ORB-v2              & 4.5   & 5.2    & 19.0      \\
\bottomrule
\end{tabular*}
\end{table}

\Cref{fig:parity_s} shows the HR factor obtained with MtS-predicted phonons versus the ground truth DFT result. The data points are colored according to the charge state of the defect. As shown in the inset box, the MAE is 0.79, which corresponds to a mean absolute percentage error (MAPE) of 12.3\%. This error may seem large. However, it should be kept in mind that the HR factor is a very sensitive quantity in general. Small changes in computational methodology, e.g., size of supercell, $k$-point mesh, or the exchange-correlation functional, can lead to variations in the predicted HR factor on the order of 10\%~\cite{alkauskas2014first, lee2025room}. For example, Alkauskas \emph{et al.}~\cite{alkauskas2014first} reported an increase in the HR factor of the NV-center in diamond from 3.27 to 3.67 (corresponding to 11\%) upon enlarging the supercell from 3$\times$3$\times$3 (216 atoms) to 4$\times$4$\times$4 (512 atoms). Similarly, the PBE xc-functional was found to yield an HR factor of 2.78 while the HSE06 xc-functional predicts 3.67, corresponding to a MAPE of 24.3\%~\cite{alkauskas2014first}.

While the overall MAE is relatively large, we note that the error increases for larger values of $S$. In fact, \cref{fig:parity_s} (a) indicates that the variance of the predictions increases with $S_{\mathrm{DFT}}$. This is quantitatively confirmed by comparing the variance of the prediction error, $|S_{\mathrm{MtS}}$ - $S_{\mathrm{DFT}}|$, for different ranges: the variance for $S_{\mathrm{DFT}} < 4$ is 0.22, while it increases to 2.41 for $S_{\mathrm{DFT}} \geq 4$. The lower variance at smaller $S$ values, see \cref{fig:parity_s} (b), is consistent with MtS providing slightly more reliable predictions in the low-$S$ regime: For the data with $S_{\mathrm{DFT}} < 4$, the MAE is only 0.25 corresponding to a MAPE of 9.25\%. This observation can be explained by noting that larger $S$-values correspond to larger displacements between the ground- and excited-state configurations. The fact that MtS predicts $S$ better for smaller values of $S$ is fortunate because many applications prefer small HR factors, making such defects more relevant.

To further assess our hypothesis that the MtS-based predictions are better for smaller displacements, we show in \cref{fig:dif_s} the deviation of the HR factor, $S_{\mathrm{MtS}}$ - $S_{\mathrm{DFT}}$, as a function of $\Delta Q$.
It clearly follows from the plot that the prediction accuracy of the MtS method tends to be higher for systems with smaller structural reorganization between the ground and excited states.
Moreover, systems with smaller structural distortions consistently show more accurate predictions, regardless of their charge state or magnetic configuration.
This finding is particularly valuable for high-throughput screening, as it suggests that MtS can reliably evaluate electron-phonon coupling in complex defect systems across different charge and spin states, provided the structural reorganization upon excitation remains moderate.
\mbox{} \\

\begin{figure}[ht!]
  \centering
  \includegraphics[width=.9\columnwidth]{img/figure2.png}
  \caption{\textbf{Difference in HR factor according to $\Delta \mathbf Q$.} The prediction accuracy is consistently high for systems with low structural reorganization between the ground and excited states.}
  \label{fig:dif_s}
\end{figure}

\par
To examine the robustness of the MtS model across different defect types, we analyzed its performance for defects in various charge states.
As seen in \cref{fig:parity_s} (a) and (b), the ML-approach based on the MtS model demonstrates nearly uniform performance across all charge states with mean absolute errors (MAE) of 0.80, 0.83, and 0.73 for charge states $q=-1, 0, +1$, respectively. The similar MAE values across different charge states demonstrate that the approach is not biased toward any particular charge configuration, providing reliable predictions for a wide range of defect systems. 

The fact that MtS performs equally well for charged and neutral defects may seem surprising, as the MtS model architecture does not take charges into account and the HR factor of different charge states of the same defect can differ significantly. 
Our explanation is that the change in charge state is mainly imprinted on $\Delta \mathrm{R}$ (and $\Delta \mathbf Q$), which is calculated using DFT. Indeed, $\Delta R$ (as well as $\Delta \mathbf Q$) can exhibit significant differences between different charge states of the same defect, with variations in $\Delta \mathbf Q$ reaching up to 89.6\%, and a mean absolute difference of  24.3\%.
On the other hand, MtS is only used to calculate the phonons in the ground state configuration. Thus, the similar performance for charged and neutral defects indicates that the phonons in the ground state are not very sensitive to the charge state of the defect. 
\mbox{} \\

\begin{figure}[ht!]
    \centering
    \includegraphics[width=0.9\linewidth]{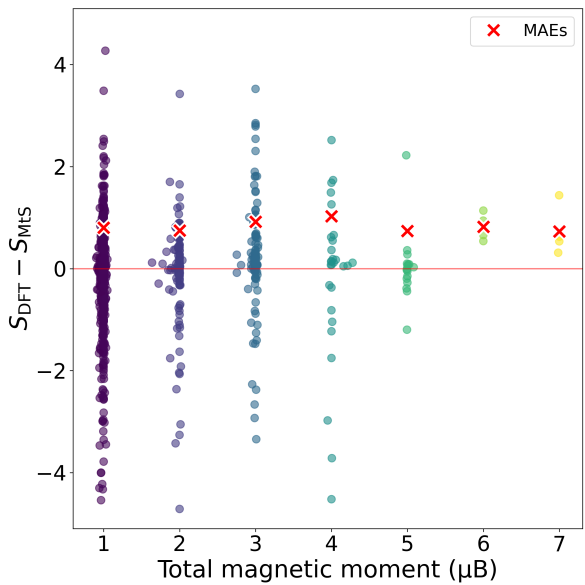}
    \caption{\textbf{The difference between the Huang-Rhys factors calculated with DFT and with MtS-predicted phonons, as a function of the total magnetic moment on the defect.} The mean absolute error (MAE) for each distribution is indicated by the red cross. The prediction accuracy of HR factors is consistent across different magnetic moments.}
    \label{fig:magmom_S}
\end{figure}

In addition to charge states, we also assessed the MtS model’s performance for defects with varying magnetic moments.
\cref{fig:magmom_S} illustrates the difference between DFT-calculated and MtS-predicted HR factors, ($S\textsubscript{DFT} - S\textsubscript{MtS}$) as a function of the magnetic moment on the defect (only magnetic defects are considered). The MAE is indicated by the red cross. 
The results with MtS-predicted phonons show consistent accuracy across the entire range of magnetic moments from 1 to 7 $\mu$B. 
This is particularly noteworthy since magnetic systems often exhibit complex electronic structure and vibrational properties due to spin-polarization effects. 
The distribution of errors remains relatively uniform across all magnetic moment values, with most predictions falling within ±2 units of the reference DFT values.

\subsection{Photoluminescence lineshape}

\begin{figure*}[p]
  \centering
  \includegraphics[width=0.95\linewidth]{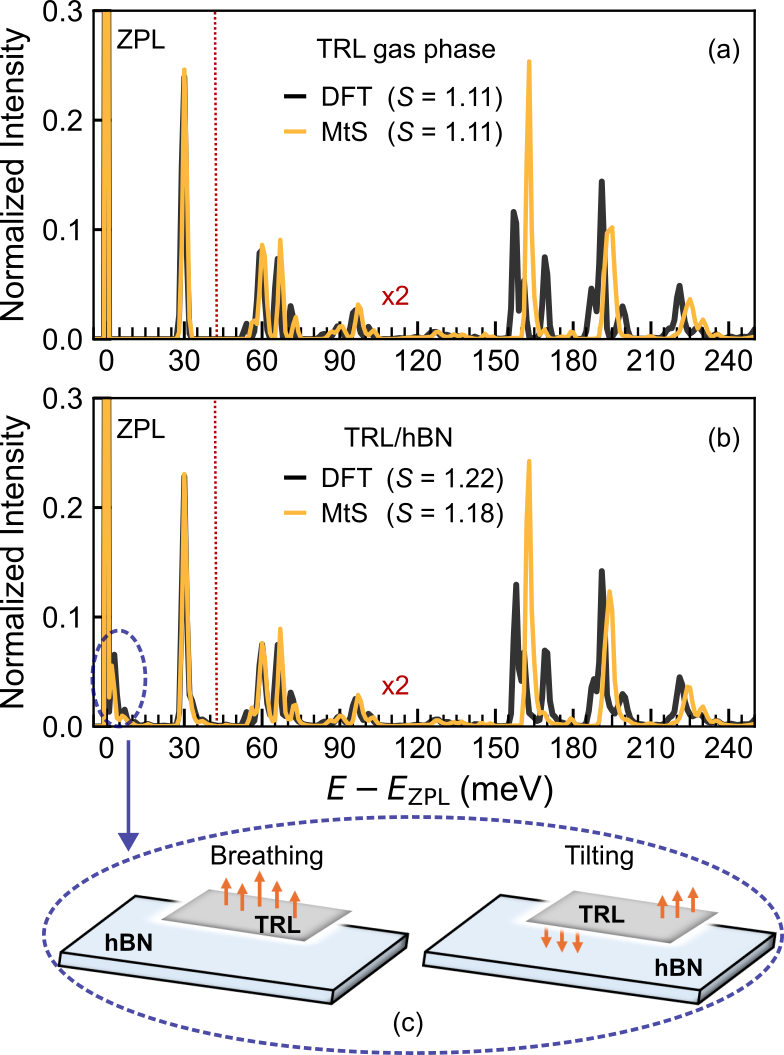}
  \caption{\textbf{Different PL spectra for multiple defect structures.} The black line is the reference spectrum (full DFT), and the orange line is obtained with MtS-predicted phonons.
  The defect systems are shown in the captions (a)–(l). It takes the form of the host crystal, followed by the point defect or the combination of two point defects occupying neighboring sites (for double defects). The letter `v' denotes vacancy defects. The charge state is shown in the superscript, and $\alpha$ refers to the majority spin channel while $\beta$ refers to the minority spin channel. The spin multiplicity is indicated at the end of each caption.}
  \label{fig:pl_plot}
\end{figure*}

For a visual assessment of the capability of our ML-framework to predict full PL spectra, we selected 12 different point defects spanning bulk and 2D host materials, one- and two-site defects, different magnetic states, and neutral, positive, and negative charge states. 

\cref{fig:pl_plot} shows the PL spectra calculated with full DFT (black line) and with MtS-predicted phonons (orange line). The first row represents three bulk defects, namely the NV$^-$ center in diamond -- a widely studied benchmark defect -- and two substitutional defects in silicon (single Cr and double Co). The last three rows represent specific cases selected from our 2D defect dataset. Each row corresponds to a specific 2D host material, with each column showing different defects within that host material. 

It is evident by visual inspection that the ML-approach performs very well in general, yielding excellent agreement with the full DFT results across all 12 defect systems.
Even in the cases of less accurate predictions, the gross features of the line shape are still well captured.

\cref{fig:worst3pred} shows PL spectra for three cases where the difference between the HR factors predicted by DFT and MtS, is large (around a factor of two). It can be seen that for these defects, the PL lineshape predicted by MtS also deviate significantly from the DFT result. In general, we find the accuracy of the HR factor to be a good descriptor for the accuracy of the PL line shape.   

The errors in the MLIP-predicted PL spectra (and HR factors) stem from two main effects. The first is the accuracy of the structure relaxation performed with the MLIP (using the DFT structure as a starting point) before computing the phonons. The second is the MLIP's inherent phonon prediction accuracy for the relevant vibrational modes (i.e., the modes with the largest $Q_k$ contribution).

We have found that the MLIP relaxation step is crucial for obtaining reliable phonons. Further, it is essential to start this relaxation from the DFT relaxed structure, as otherwise the MLIP can lead to incorrect defect geometries. This deficiency could stem from an under-representation of point defect structures in the proprietary MtS training data.

\subsection{Hybrid phonons approach}
To address the limitations of MLIPs in accurately predicting the HR factor and PL spectra, we developed a hybrid phonon approach that combines the computational efficiency of ML with the accuracy of DFT in the critical region, i.e., the neighborhood around the defect site.

In the hybrid approach, the force constant matrix, $\Phi_{ij}^{\alpha\beta}$, is constructed using DFT forces (on all atoms of the supercell) upon displacement of atoms in a certain cut off radius, $r_c$, of the defect center, $\mathbf{r}_{\mathrm{D}}$. MLIP forces are used upon the displacement of all remaining atoms in the supercell. The total number of DFT calculations required is thus reduced by a factor $N_c/N$, where $N$ is the total number of atoms in the supercell and $N_c$ is the number of atoms within the cutoff radius.  Precisely, the (symmetric) force constant matrix is given by 
\begin{equation}
    \Phi_{ij}^{\alpha\beta,\text{hybrid}} = \begin{cases}
    -\dfrac{\partial F_i^{\alpha,\text{DFT}}}{\partial u_j^\beta} & 
    \begin{aligned}
        &\text{if}\;|\mathbf{r}_j - \mathbf{r}_{D}| \leq r_c \\
    \end{aligned} \\[12pt]
    -\dfrac{\partial F_i^{\alpha,\text{ML}}}{\partial u_j^\beta} & \text{otherwise}
\end{cases}
\end{equation} 
The dynamical matrix obtained from the hybrid force constant matrix inherits its spatial selectivity, ensuring that phonon modes with significant amplitude near the defect are governed by DFT accuracy.


Fig. \ref{fig:hybridPL} shows the hybrid approach for defects in (a) h-BN, (b) WSe\textsubscript{2}, and (c) MoTe\textsubscript{2} -- all with large errors in predicting the HR factor and PL lineshape.
The hybrid phonons with increasing cutoff radius $r_c$ systematically converge toward the DFT result, demonstrating effective correction of ML predictions. 
We tested this hybrid approach on a subset of the dataset comprising the 50 structures with the largest error in the HR factor and $S_{\mathrm{DFT}}<10$.
Notably, using $N_c$ in the range of 16--20 atoms, corresponding to $r_c$ = 4--5~\AA, substantially improves both the HR factor and the PL spectra, reducing the average error on $S$ from 48.7\% to 5.1\%. This indicates that the forces predicted by MtS on atoms in the vicinity of the defect are responsible for the majority of the prediction error, which can be effectively mitigated by enforcing DFT accuracy locally around the defect.

\subsection{Molecular emitter}

In addition to point defects, molecules immobilized on surfaces, such as hexagonal boron nitride (hBN), have recently emerged as promising solid-state single-photon emitters, providing complementary advantages to defect-based color centers. These systems exhibit highly localized electronic transitions, narrow emission linewidths, and chemical tunability, making them particularly attractive for quantum optics applications. To extend our benchmark beyond point defects, we consider the case of a molecular emitter physisorbed on a 2D substrate. Specifically, motivated by recent experimental and theoretical work~\cite{smit2023sharp, wang2025two}, we focus our attention on the terrylene molecule (TRL) adsorbed on a hBN surface. This system has been found to emit single photons efficiently within a narrow energy range around 2.1 eV.  

We evaluate the performance of the MtS approach by calculating the HR factor and PL spectrum of the isolated TRL in the gas phase and TRL/hBN interface using full DFT and the MtS-based approach. To capture the crucial van der Waals forces between the TRL molecule and hBN substrate, the DFT calculations were performed using the PBE xc-functional with the D3 dispersion correction. For consistency, the D3 correction was also applied to the MtS energies and forces. For the isolated TRL, both DFT and MtS yield identical HR factors of 1.11. Moreover, as can be seen in \cref{fig:interface_hbn}(a), the PL spectra predicted by the two methods are in excellent agreement with only small deviations for the sidebands corresponding to the most energetic molecular vibrations in the range 150-210 meV. In fact, MtS fails to predict the multi-sub-peak nature of these sidebands and instead predicts single peaks. Nevertheless, it is clear that the MtS-approach allows us to predict peak positions and relative intensities with high accuracy. In the TRL/hBN interface, the HR factors predicted by DFT (1.22) and MtS (1.18) are also in good agreement, and the resulting PL spectra, shown in \cref{fig:interface_hbn}(b), remain closely aligned. This highlights the robustness of the MtS approach, even for hybrid molecular–substrate systems.

To further interpret the spectral features, we analyze the vibrational origin of the sidebands. As shown in \cref{fig:interface_hbn}, the sidebands above 25 meV are nearly identical for both the isolated and adsorbed molecule, suggesting that these features are intramolecular in origin and largely unaffected by the presence of the substrate. The main difference between the two cases lies in the appearance of two low-energy sidebands around 5 meV in the TRL/hBN system. These features arise from a breathing and tilting mode of the TRL molecule that couples weakly to the hBN surface, as illustrated in \cref{fig:interface_hbn}(c). Notably, this mode is well captured by both DFT and MtS, further supporting the accuracy of our approach. A detailed normal mode analysis, including the role of intermolecular and substrate interactions, can be found in our previous work~\cite{wang2025two}. In brief, all vibrational modes contributing to the sidebands (except for the low-frequency ones) originate from the molecule itself, and the hBN substrate plays only a minor perturbative role, consistent with the earlier finding~\cite{smit2023sharp}.

\begin{figure*}[ht!]
    \centering
    \includegraphics[width=\linewidth]{img/figure5.pdf}
    \caption{\textbf{PL spectra for three hand-picked cases with large deviation in HR factor prediction.} The defects include - a double defect in BN, a extrinsic substitution in WSe\textsubscript{2} and a vacancy in MoTe\textsubscript{2}. The spectra predicted by MtS differ in peak positions and the intensity.}
    \label{fig:worst3pred}
\end{figure*}

\vspace{30pt}
\begin{figure*}[htbp]
    \centering
    \includegraphics[width=\linewidth]{img/figure6.pdf}
    \caption{\textbf{PL spectra constructed using the hybrid approach.} PL spectra are shown for three cases with large deviation in HR factor prediction, with varying cut-off radius $r_c$ in the hybrid approach . $N_c$ represents the number of atoms using DFT force constants within radius $r_c$. As $r_c$ is increased, the PL spectra systematically converges towards the DFT result. $N_c$ in the range of 16-20 atoms around the defect (corresponding to $r_c$ = 4-5 \AA), substantially improves both the PL lineshape and HR factor.}
    \label{fig:hybridPL}
\end{figure*}

\begin{figure}[t]
  \centering
  \includegraphics[width=.9\columnwidth]{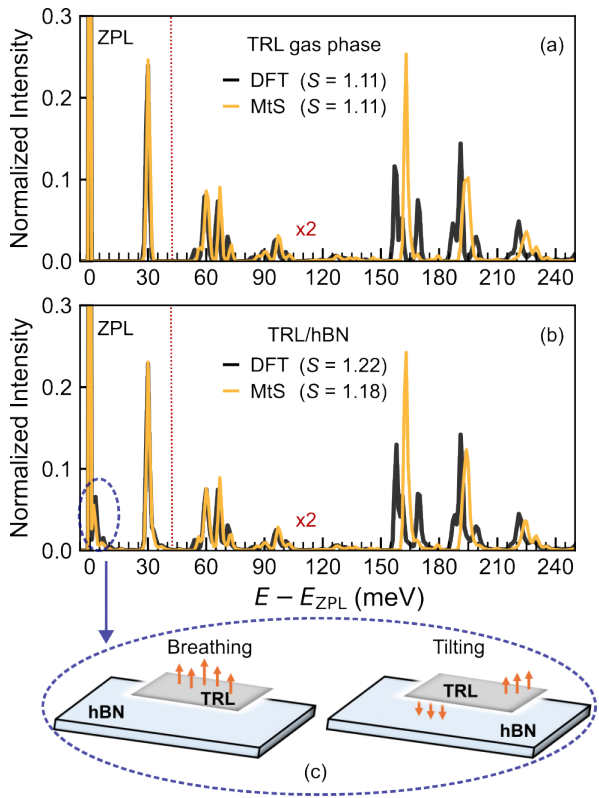}
  \caption{\textbf{A comparison of the photoluminescence spectra obtained from DFT (black curve) and MtS (yellow curve) for the (a) TRL in the gas phase and for (b) TRL adsorbed on hBN.} 
  The low-energy side bands below 10 meV in panel (b) correspond to a breathing and tilting mode of the adsorbed molecule, and are not present in the gas phase spectrum. All spectra are normalized, with the intensity of the ZPL set to unity and the ZPL positions shifted to 0. The spectral region beyond 45 meV is enhanced by a factor of 2 for clarity. (c) Vibrational modes corresponding to the peaks highlighted by the blue circle in (b).}
  \label{fig:interface_hbn}
\end{figure}

\section{Discussion}
\label{sec:conclusion}

In summary, we combined state-of-the-art universal MLIPs for phonon prediction with DFT excited state calculations for accelerated determination of optical properties of point defects. Benchmarking seven different MLIPs on a dataset comprising 791 point defect PL spectra and Huang-Rhys factors identified the Mattersim-V1-5M (MtS) as the best model for the present application. For a representative set of 12 point defects, we calculated the full PL lineshape and found excellent agreement between full DFT and the MtS-accelerated approach, with the latter being faster by an order of magnitude.  

To further test the versatility of the MtS-based method, we conducted a case study of a molecular emitter adsorbed on a substrate. Also, for this system, MtS performed very well, reproducing both quantitative and qualitative aspects of the DFT spectrum.

MtS demonstrates adequate performance across diverse defect systems regardless of their charge state or magnetic properties. This consistency suggests that neither charge state nor magnetic properties significantly impact the prediction accuracy of our ML-accelerated approach. This may seem surprising since the MLIP is neither aware of charges nor magnetic moments. We proposed that the consistently good performance is due to the fact that the MLIP is only used to compute the phonons in the ground state configuration, while the displacement vector, $\Delta \mathbf R$, which is more sensitive to the charge/magnetic state of the defect, is calculated using DFT. 

We believe that this work establishes a strong foundation for significantly accelerating the prediction of optical properties and PL spectra of atomic and molecular emitters by leveraging the efficiency of universal MLIPs for phonon predictions. These advances hold promise to boost high-throughput investigations and expand the scope of computationally accessible color center systems.
\mbox{}\\

\section{Methods}
\label{sec:method}

\subsection{Ab initio GS and ES calculations}
All spin-polarized DFT calculations were performed using the GPAW electronic structure code~\cite{Mortensen2024}, employing the Perdew-Burke-Ernzerhof (PBE) exchange-correlation functional~\cite{Perdew1996}. 
A plane-wave cutoff of 800~eV and a $k$-point density of 3~\AA~ were used for structural relaxations and electronic ground-state calculations. Fermi smearing of 0.02~eV was applied in all calculations.

For excited-state calculations, we employed GPAW's implementation of the direct-optimization of maximally overlapping orbitals method (DO-MOM) ~\cite{jlevi2021}, which promotes a single electron from the HOMO to LUMO while maintaining maximum overlap with ground-state orbitals.
The DO-MOM algorithm uses a nested optimization approach: an inner loop finds stationary points in the subspace of occupied and virtual orbitals through unitary transformations, while an outer loop minimizes an auxiliary energy functional in the full orbital space. Excitations are performed separately for $\alpha$ (spin-up) and $\beta$ (spin-down) channels. Excited states preserve the total spin; for example, a triplet ground state ($S=1$) yields triplet excited states through same-spin excitations within each spin channel.
The excited state and DFT phonon calculations were performed at the $\Gamma$-point.

\subsection{MLIP phonons}
For universal MLIP phonon calculations, we used the model trained in Ref.~\cite{mattersim}, specifically version Mattersim-V1. Before computing phonons, we performed the geometry relaxation with the MLIP, starting from the DFT ground state geometry. This is done to ensure that the structure is in the minima of the MLIP energy surface. The structure relaxation was done using the fast inertial relaxation engine (FIRE)~\cite{Bitzek2006}, with a force convergence criterion of 0.005 eV/\AA. The phonons were calculated using finite-displacement methods available through phonopy~\cite{togo_distributions_2015,togo_first-principles_2023}.

\subsection{Approximations}
Our approach relies on three key approximations: 
\begin{enumerate}[(i)]
    \item The Born-Oppenheimer approximation, which is justified since the host materials are semiconductors with large band gaps ($>$1 eV), ensuring that non-adiabatic coupling between electronic and nuclear degrees of freedom remains negligible during optical transitions.
    \item Single-determinant excited states calculations via DO-MOM, neglecting explicit many-body correlation effects.
    \item The potential energy surface around the ground state and excited state configurations are assumed to be harmonic and identical up to the displacement ($\Delta\mathbf{R}$). In other words, the vibrational energies and normal modes in the ground state and excited states are assumed to be identical, and anharmonic effects are ignored. 
\end{enumerate}

Partial Huang-Rhys factors are calculated using DFT-derived structural displacements ($\Delta\mathbf{R}$) between relaxed ground and excited state geometries, projected onto harmonic phonon modes for the ground state. This assumes that phonon modes are similar between ground and excited states and neglects anharmonic effects in the vibrational potential surfaces.

\section{Data availability}
\label{sec:data_ava}
The point defect database is available in the CMR repository under Quantum Point Defects Database \textsc{(QPOD)}, which can be accessed from \url{https://qpod.fysik.dtu.dk/}. 
The benchmark data used in this manuscript is available in QPOD under the "biomag2d" structure origin label.

\section{Code availability}
\label{sec:code_ava}
The code used in this work is based on the \texttt{asr-lib} library, branch \textsc{2D-screening-20240821}, which can be accessed from \url{https://gitlab.com/asr-dev/asr-lib/-/tree/2D-screening-20240821/defects?ref_type=heads}.

\section{Acknowledgements}
The authors acknowledge funding from the Horizon Europe MSCA Doctoral network grant n.101073486, EUSpecLab, funded by the European Union, and from the Novo Nordisk Foundation Data Science Research Infrastructure 2022 Grant:  A high-performance computing infrastructure for data-driven research on sustainable energy materials, Grant no. NNF22OC0078009.
F.N. has received funding from the European Union’s Horizon 2020 research and innovation program under the Marie Skłodowska-Curie Grant Agreement No. 899987.
K. S. T. is a Villum Investigator supported by VILLUM FONDEN (grant no. 37789).

\section{Author contributions}
A.L. and K.S. developed the initial concepts and wrote the main manuscript. F.N., K.S., and H.W. developed the workflow for screening defects for PL spectra calculations using DFT and performed DFT calculations for the 2D defects dataset of PL spectra. M.J. performed DFT calculations for bulk defects corresponding to Fig. 4 (a-c).  A.L. benchmarked the uMLIPs for the application, prepared Table 1, and contributed data corresponding to the MLIP part. K.S. prepared figures 1-6, and H.W. prepared Fig. 7 and contributed to the molecular emitters section. K.S.T. supervised the project and edited the manuscript. All authors reviewed the manuscript.

\section{Competing interests}
The authors declare no competing interests.

\bibliography{bib.bib}

\end{document}